# DEVELOPMENT OF THE CHARGE PARTICLE DETECTOR BASED ON CVD - DIAMOND

*S.V. Akulinichev, V.S. Klenov, L.V. Kravchuk, S.G. Lebedev, A.V. Feschenko, V.E. Yants*
*Institute for Nuclear Research of the Russian Academy of Sciences, 60-th October Anniversary Prospect, 7A, Moscow, 117312*
E-mail: *klenov@inr.ru*

High radiation hardness, chemical resistance, high temperature operation capabilities stimulate a growing interest to use diamond materials as detectors of ionizing radiation. Samples of CVD-diamond materials in sizes 4x3 mm and 4×1 mm with thickness from 50 microns up to 500 microns have been grown in INR RAS using a DC glow discharge in a mixture of gases CH4/H2 on molybdenum substrates.

PACS: 29.40.Wk; 81.05.T

## 1. INTRODUCTION

A number of unique properties of diamond such as extremely high radiation hardness, chemical resistance against all chemicals, absolute non-toxicity call for an increasing interest to use diamond materials as detectors of ionizing radiation operating in hostile environments or in conditions, imposing special requirements to stability of measurement of a doze, for example in medical installations for radiotherapy. Moreover, the atomic number of diamond Z = 6 that is close to the effective atomic number of a soft tissue Z = 7.4, so the diamond is a nearly tissue equivalent, that allows to avoid energy dependent corrections of the detector signal. Initially, natural diamonds with suitable electronic properties were used in radiation detection [1]. The main disadvantage of the natural diamond detectors is a high cost due to extremely rare detector-grade natural diamond (Type IIa), which limits the availability of these detectors and moreover, electronic properties of diamond stones within the Type IIa category can vary strongly. Therefore, the production of sufficiently cheap diamond plates with sizes at least 4 mm and thickness 50 –500 μm with sufficient quality to build the detector is rather urgent problem.

The promising technology for synthesis of diamond materials is the Chemical Vapour Deposition (CVD) technology, which allows to grow diamond material plates in controllable vacuum with specified thickness and sizes, which are determined by sizes of substrates and duration of the process. However, the CVD-diamond has a polycrystalline structure with crystallites sizes about 10-20 % of thickness of grown plates, and crystallites bounders could act as the traps and decrease the charge collection efficiency [2,3].

## 2. APPARATUS FOR CVD-DIAMOND PLATES SYNTHESIS

We have developed a CVD apparatus based on DC glow discharge for manufacture of cost effective CVD-diamond plates. This apparatus is schematically shown in Fig.1. In the reaction chamber glow discharge is sustained in a mixture of gases $CH_4$ and $H_2$ between molybdenum cathode 25 mm diameter and molybdenum anode 11 mm diameter. The cathode is mounted on the copper water-cooled holder. The front surface of the anode is polished and segmented by grooves 0.5 mm in depth into sites $4\times1$ $mm^2$ or $4\times3$ $mm^2$, which simultaneously are the substrates for growth of diamond plates. The conditions of synthesis (gas pressure, power density in the discharge) are fitted in such a manner, that the growth of diamond takes place only on a surface of substrates.

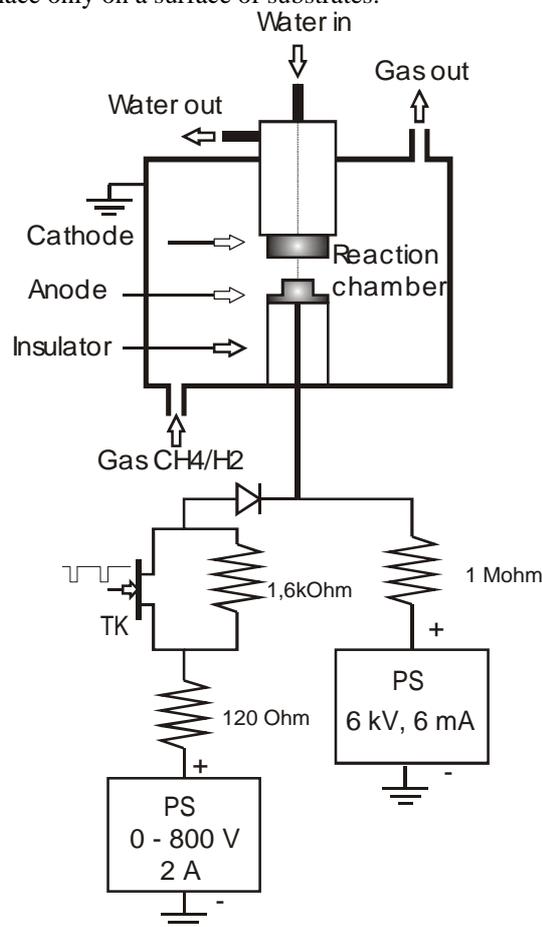

*Fig.1. Scheme of CVD-diamond plates synthesis*

It is well known that glow discharge with a current close to a critical one has a considerable probability to transform into an arc discharge with drastic contraction of a discharge channel, which could result in damages of growing material. To decrease a probability of these transitions, similarly to [4], the pulse operating mode of the discharge is used. The duration of pulses and pauses of a discharge current is set by a transistor modulator TK. The resistor connected in parallel to TK is important as it provides decreasing of a current in a pause not to zero, but up to a magnitude ~$0.15I_{nom}$, which facilitates the subsequent transition to the rated current and rises stability of operation. Furthermore the transistor modulator TK provides fast (~ 10 μs) switching-off a discharge current source in case the discharge starts to transform into an arc mode and a current exceeds a preset value. The additional power supply with a constant voltage of 6 kV is connected to the discharge gap in series with 1 MOhm resistor for providing a discharge ignition in case of accidental extinction.

Synthesis of CVD-diamond plates was carried out under the following conditions in reaction chamber: typical gas mixture – 2.5%$CH_4$ in $H_2$, gas pressure 300 Torr, discharge voltage 540 V, discharge current 1.7 A, current duty factor 95%.

Oscillogram of discharge current is shown on Fig.2. Material growth rate in these conditions was around 12÷15 μm/hour.

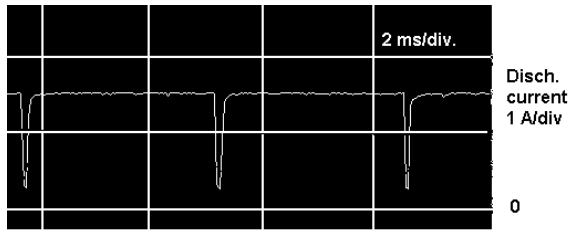

*Fig.2. Oscillogram of the discharge current*

## 3. CHARACTERIZATION OF THE GROWN MATERIAL

In deposition runs lasted from 3 to 35 hours CVD-diamond plates with the sizes 4×1 mm$^2$ and 4×3 mm$^2$ and the thickness from 50 to 500 μm have been grown. After deposition and cooling down the CVD-diamond plates can be easily detached from the substrates due to different thermal expansion of diamond and molybdenum. The grown material has a polycrystalline structure with a clearly visible in the cracked samples crystallites of columnar shape elongated along the growth direction. The facets of about 0.1 ÷ 0.2 plate thickness can be observed on the crystallites at a growth side of the plate. Fig.3 shows the image of growth side of plate with thickness 100 μm (scanning electron microscope). Fig.4 shows the spectra of Raman scattering analysis, which was performed with argon laser on the wavelength λ =514 nm, narrow diamond peak at 1333 cm$^{-1}$ is clear visible.

X-ray diffraction analysis of the plates was made using the diffractometer DRON-3 with CuKα1 line with λ = 0.154057 nm. Fig.5 shows a XRD pattern measured at the growth side of 500 μm plate.

## 4. CVD-DIAMOND DETECTION PERFORMANCE

The surface morphology of plates from a growth side and from a substrate side is essentially different. Whereas on the growth side the typical sizes of crystallites make up tens of micron, on the substrate side these sizes do not exceed a micron. The numerous defects on crystallites borders can serve as traps for the charges induced in the detector by ionizing particles.

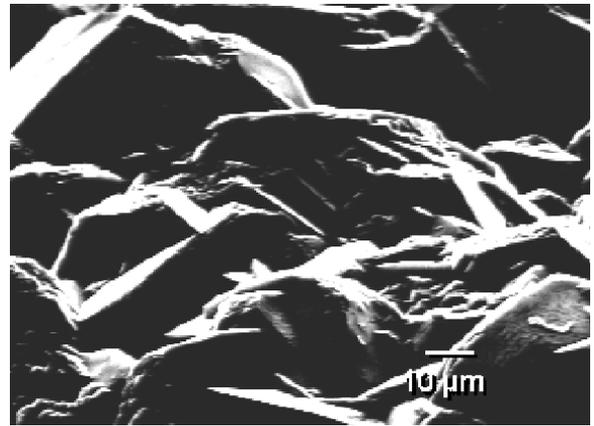

*Fig.3. Scanning electron microscope (SEM) image of growth side 100 μm thickness plate*

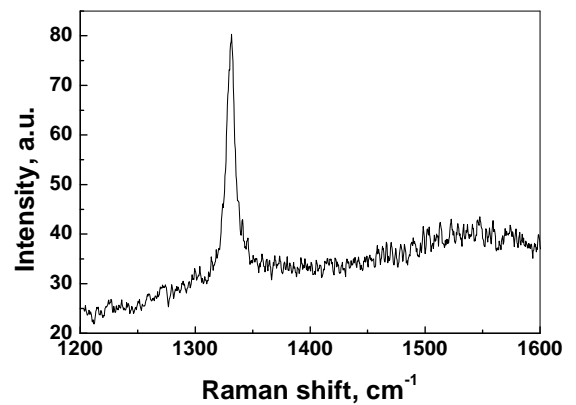

*Fig.4. Raman spectra at growth surface of CVD diamond plate*

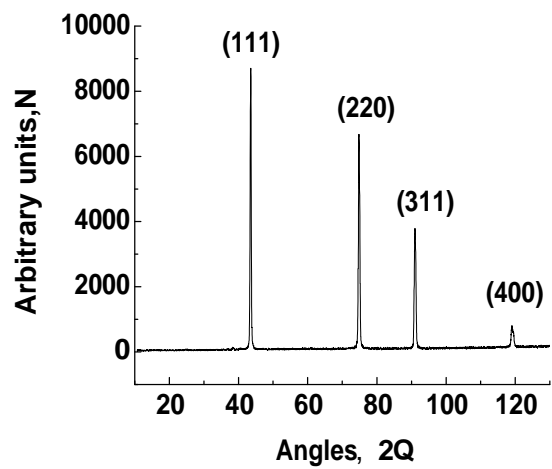

*Fig.5. XRD pattern for 500 μm plate*

In order to decrease an influence of these traps the coplanar type detecting device [5] was made with both electrodes located on the growth side of the plate. The distance between the electrodes makes up 200 micron. For measurement of efficiency of collection of the charges induced by incident alpha-particles, the installation schematically represented on Fig.6 has been assembled.

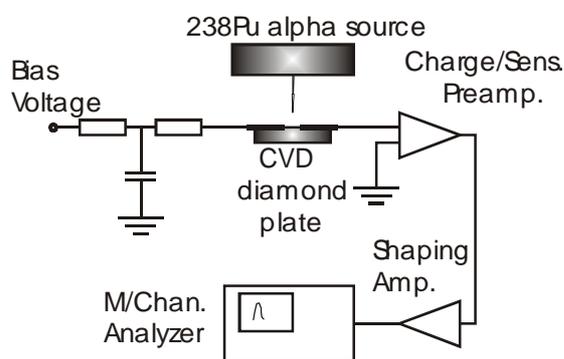

*Fig.6. A schematic diagram of the charge collection efficiency measurement.*

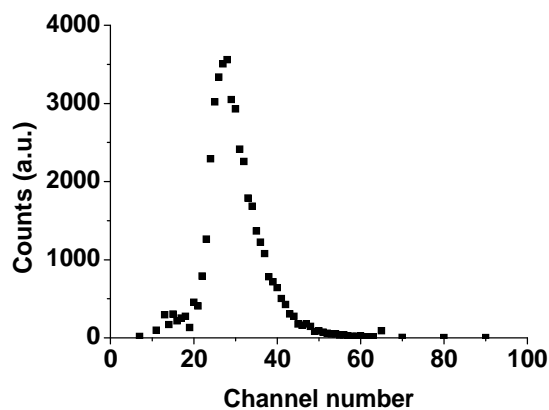

*Fig.7. Pulse height spectra for bias voltage 40 V*

We used a $^{238}$Pu α-particles source which emits α-particles with the energy $E\alpha = 5.5$ MeV.

The estimated range of this particles in a diamond is ~ 13 μm. The total charge $Q_{ind}$, induced by α-particle in a diamond $Q_{ind} = eE_\alpha/\varepsilon$, where $\varepsilon = 13$ eV is the energy of electron-hole creation in diamond. The charge collected by non-uniform inter-electrode field of bias voltage feed at entrance of charge sensitive preamplifier followed by a shaping amplifier (Schlumberger Type 7129) and multichannel analyzer (Norland 5300). Pulse height specters were measured as differences of counts with and without α-source for exception of electronic noise of a system. Pulse height spectra for a bias voltage of 40 V is shown on Fig.7. The charge collection efficiency was estimated as a ratio of collected and induced charges and comes to around 1%.

## 5. CONCLUSION

Our measurements have demonstrated that relatively cheap CVD diamond, produced by glow discharge, is suitable for detecting of charged particles. The next tack which will be studied is the stability of such diamond detector against neutron and particle irradiation which can be done at the RADEX facility [6-9] of INR RAS. Further investigations of radiation hardness and stability should be made in the radiological center of INR RAS on the beams of 200 MeV protons and 6 MeV photons. It is believed, that CVD diamond based charge particle detectors will continue the sucsessful sequence of particle detectors developed up today in INR RAS [10-13].

## ACKNOWLEGMENTS
Special thanks are given to I.I. Vlasov (GPI RAS) for performing of Raman analysis and V. Vlasenko (CryoLab, MSU) for SEM images.